\documentclass[%
 reprint,
showpacs,preprintnumbers,
 amsmath,amssymb,
 aps,
prb,
]{revtex4-1}

\usepackage{graphicx}
\usepackage{dcolumn}
\usepackage{bm}% bold math
\usepackage{color}
\usepackage{ulem}

\begin{document}

\preprint{APS/123-QED}

\title{
Topological 
semimetal-to-insulator 
phase transition between noncollinear and noncoplanar multiple-$Q$ states 
on a square-to-triangular lattice
}

\author{Satoru Hayami and Yukitoshi Motome}
\affiliation{%
 Department of Applied Physics, University of Tokyo, Tokyo 113-8656, Japan 
 }%

\begin{abstract}
Noncollinear and noncoplanar magnetic orders lead to unusual electronic structures and transport properties. 
We here investigate two types of multiple-$Q$ magnetically ordered states and a topological phase transition between them in two dimensions.  
One is a coplanar but noncollinear double-$Q$ state on a square lattice, which is a semimetal accommodating massless Dirac electrons. 
The other is a noncoplanar triple-$Q$ state on a triangular lattice, which is a Chern insulator showing the quantum anomalous Hall effect. 
We discuss the peculiar electronic structures in these two multiple-$Q$ states in a unified way on the basis of the Kondo lattice model, which suggests a quantum phase transition between the two states in a continuous change of lattice geometry between the square and triangular lattices. 
We systematically examine the possibility of such a transition by using the mean-field approximation for the ground state of the periodic Anderson model. 
After clarifying the parameter region in which the double-$Q$ (triple-$Q$) state is stabilized on the square (triangular) lattice, 
we show that a continuous topological phase transition indeed takes place between the double-$Q$ Dirac semimetal and the triple-$Q$ Chern insulator on the square-to-triangular lattice. 
The nature of the transition is discussed by the topologically-protected edge states as well as the bulk magnetic and electronic properties. 
The results indicate that unusual critical phenomena may occur at finite temperature related with multiple-$Q$ chiral spin-liquid states. 
\end{abstract}
\maketitle

\section{Introduction}
\label{sec:Introduction}

The search for topologically nontrivial states has attracted growing interest in condensed matter physics~\cite{Hasan_RevModPhys.82.3045,Qi_RevModPhys.83.1057}. 
Among them, a topological insulator is a new quantum state of matter~\cite{Haldane_PhysRevLett.61.2015,Kane_PhysRevLett.95.146802,Kane_PhysRevLett.95.226801,Murakami_PhysRevLett.97.236805,bernevig2006quantum}. 
In this state, although the bulk of the system is insulating with an energy gap opened by the spin-orbit coupling, a spin current flows along the edge/surface of the system. 
The current flow is dissipationless and gives rise to a quantized spin Hall response. 
Such a quantum spin Hall effect has been experimentally observed in two-dimensional quantum wells, such as HgTe/CdTe~\cite{konig2007quantum} and InAs/GaSb~\cite{liu2008quantum,knez2011evidence}, and in three-dimensional bulk systems, such as Bi$_2$Se$_3$, Bi$_2$Te$_3$, and Sb$_2$Te$_3$~\cite{hsieh2008topological,zhang2009topological,xia2009observation,chen2009experimental}. 

On the other hand, topologically nontrivial states are also realized in magnetically-ordered systems with broken time reversal symmetry. 
Such topological insulators with magnetic ordering are called magnetic Chern insulators, as they are classified according to nonzero values of the Chern number~\cite{Thouless_PhysRevLett.49.405}. 
Remarkably, the magnetic Chern insulators exhibit a quantum anomalous (topological) Hall effect even in the absence of an external magnetic field. 
The origin of the anomalous Hall effects has been discussed in terms of the spin scalar chirality and associated Berry phase in spin space~\cite{berry1984quantal,Loss_PhysRevB.45.13544,Ye_PhysRevLett.83.3737,Ohgushi:PhysRevB.62.R6065}.  

Such noncoplanar scalar chiral orders have been theoretically explored in the Kondo lattice model on several frustrated lattice structures: 
a face-centered-cubic~\cite{Shindou_PhysRevLett.87.116801}, triangular~\cite{Martin_PhysRevLett.101.156402,Akagi_JPSJ.79.083711,Kato:PhysRevLett.105.266405,Akagi_PhysRevLett.108.096401,Barros_PhysRevB.88.235101,Rahmani_PhysRevX.3.031008,Akagi_doi:10.7566/JPSJ.82.123709,Ozawa_doi:10.7566/JPSJ.83.073706}, checkerboard~\cite{Venderbos:PhysRevLett.109.166405}, pyrochlore~\cite{Chern:PhysRevLett.105.226403}, and kagome lattices~\cite{Ohgushi:PhysRevB.62.R6065,Ishizuka_PhysRevB.87.081105,chern2012quantum,barros2014novel,ghosh2014phase}. 
These chiral ordered states will be important for potential applications to spintronics, as the topological nature can be controlled by external magnetic field and pressure as well as temperature. 
Furthermore, they are beneficial owing to the possibility of acquiring a wide energy gap determined by the exchange interaction rather than the spin-orbit coupling. 
To stimulate experimental exploration of such exotic states, it is desired to systematically study how and when the noncoplanar magnetic orders become stable. 
It is also interesting to study how the topological nature changes in 
magnetic phase transitions. 

In the present study, we investigate the stability of magnetic Chern insulators and their phase transitions, with focusing on multiple-$Q$ magnetically ordered states.
The multiple-$Q$ states are characterized by more than one ordering wave vectors, and tend to accommodate noncollinear and noncoplanar magnetic orders. 
Hence, the multiple-$Q$ states in itinerant electron systems provide an ideal prototype for the magnetic Chern insulators. 
In fact, some of the above-mentioned states in the Kondo lattice model are the multiple-$Q$ Chern insulators~\cite{Shindou_PhysRevLett.87.116801,Ohgushi:PhysRevB.62.R6065,Martin_PhysRevLett.101.156402,Akagi_JPSJ.79.083711,Venderbos:PhysRevLett.109.166405,barros2014novel}. 

Specifically, we here focus on a triple-$Q$ state on a triangular lattice and a double-$Q$ state on a square lattice. 
The former exhibits a ferroic order of the spin scalar chirality associated with four-sublattice noncoplanar magnetic order, leading to a magnetic Chern insulating state~\cite{Martin_PhysRevLett.101.156402,Akagi_JPSJ.79.083711}. 
Meanwhile, the latter shows a coplanar but noncollinear four-sublattice order, resulting in a semimetallic state with massless Dirac electrons~\cite{Yamanaka_PhysRevLett.81.5604,Agterberg_PhysRevB.62.13816,Chen_PhysRevB.81.064420}. 
First, we discuss the topologically-nontrivial electronic structures for these two states in the Kondo lattice model. 
We show that a topological phase transition between the two states can occur while changing the lattice geometry. 
Next, we discuss the stability of the multiple-$Q$ states and the possibility of such a topological phase transition by mean-field calculations for the periodic Anderson model. 
We show that the double-$Q$ and triple-$Q$ states are robustly realized on the square and triangular lattices, respectively, in a wide range of parameters. 
Furthermore, we find that the model indeed exhibits a continuous topological phase transition between the two states by changing the lattice geometry. 
We discuss the detailed nature of the transition by presenting the change of the topologically-protected gapless edge states as well as the bulk physical properties. 

The organization of this paper is as follows. 
In Sec.~\ref{sec:Multiple-Q states in spin-charge coupled systems}, after introducing the double-$Q$ and triple-$Q$ states on the square and triangular lattices, we discuss the peculiar electronic structures for bulk and edge states and an anticipated quantum phase transition between them by using the Kondo lattice Hamiltonian. 
In Sec.~\ref{sec:Stability of multiple-Q states}, by using the mean-field approximation for the periodic Anderson model, we show that the system exhibits a continuous quantum phase transition between the double-$Q$ Dirac semimetal and the triple-$Q$ Chern insulator. 
Section~\ref{sec:Summary} is devoted to a summary and concluding remarks. 

\section{Multiple-$Q$ states and their electronic structure}
\label{sec:Multiple-Q states in spin-charge coupled systems}

In this section, we examine the fundamental magnetic and electronic nature of the multiple-$Q$ states that we study in this paper. 
In Sec.~\ref{sec:Multiple-Q states on lattices}, we introduce the definition of multiple-$Q$ magnetic orders. 
In Sec.~\ref{sec:Kondo Hamiltonian}, we present the Kondo lattice Hamiltonian for the double-$Q$ and triple-$Q$ states in a unified way. 
We show their electronic structures for both bulk and peculiar gapless edge states in Sec.~\ref{sec:Electronic structure}. 
In Sec.~\ref{sec:Topological phase transition between square and triangular lattices}, we discuss an anticipated phase transition between these two states by modifying the lattice geometry. 

\subsection{Multiple-$Q$ states}
\label{sec:Multiple-Q states on lattices}

\begin{figure}[hbt!]
\begin{center}
\includegraphics[width=1.0 \hsize]{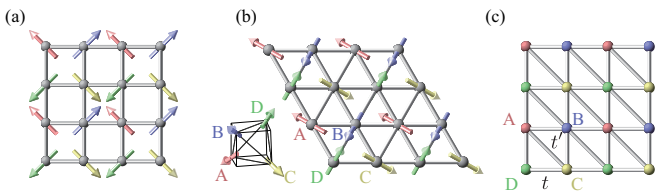} 
\caption{
\label{Fig:multipleQ_ponti_squ_tri}
(Color online) Schematic pictures of (a) a double-$Q$ state with a coplanar spin configuration on a square lattice and (b) a triple-$Q$ state with a noncoplanar spin configuration on a triangular lattice. 
The arrows denote the directions of local magnetic moments. 
The inset of (b) shows the directions of magnetic moments at the four-sublattice sites in the triple-$Q$ state. 
(c) The square lattice with diagonal bonds, which 
is topologically equivalent to the triangular lattice in (b). 
$t$ and $t'$ represent the hopping amplitudes in the models in Eqs.~(\ref{eq:Kondo_lattice_Ham}) and (\ref{eq:PAM_Ham}). 
We call this the square-to-triangular lattice. 
}
\end{center}
\end{figure}

In this study, we consider double-$Q$ and triple-$Q$ states on square, triangular, and their intermediate lattices. 
The order parameter for the double-$Q$ state is written by 
\begin{align}
\label{eq:doubleQ_definition}
\langle \bm{S}_{i} \rangle \propto 
(\cos \bm{Q}_1 \cdot \bm{r}_i, \cos \bm{Q}_2 \cdot \bm{r}_i, 0), 
\end{align}
while that for the triple-$Q$ state is given by
\begin{align}
\label{eq:tripleQ_definition}
\langle \bm{S}_{i} \rangle \propto 
(\cos \bm{Q}_1 \cdot \bm{r}_i, \cos \bm{Q}_2 \cdot \bm{r}_i, \cos \bm{Q}_3 \cdot \bm{r}_i). 
\end{align}
Here, $\bm{Q}_1$, $\bm{Q}_2$, and $\bm{Q}_3$ stand for the wave vectors characterizing the multiple-$Q$ states; $\bm{r}_i$ is the position vector of the site $i$. 
Equation~(\ref{eq:doubleQ_definition}) defines a noncollinear but coplanar spin configuration as exemplified on the square lattice in Fig.~\ref{Fig:multipleQ_ponti_squ_tri}(a)~\cite{Agterberg_PhysRevB.62.13816}, while Eq.~(\ref{eq:tripleQ_definition}) gives a noncoplanar configuration as illustrated on the triangular lattice in Fig.~\ref{Fig:multipleQ_ponti_squ_tri}(b)~\cite{Martin_PhysRevLett.101.156402,Akagi_JPSJ.79.083711}; 
both of them are represented by superpositions of single-$Q$ states. 

In the following, we treat these two multiple-$Q$ states in a unified way by regarding the triangular lattice as a topologically equivalent square lattice with diagonal bonds, as shown in Fig.~\ref{Fig:multipleQ_ponti_squ_tri}(c). 
We call this structure the square-to-triangular lattice. 
Then, the double-$Q$ state is given by the superposition of $(\cos \pi r_i^x, 0, 0)$ and $(0, \cos \pi r_i^y, 0)$, which is described by taking $\bm{Q}_1=(\pi, 0)$ and $\bm{Q}_2=(0, \pi)$ in Eq.~(\ref{eq:doubleQ_definition}). 
Here, $\bm{r}_i=(r_i^x, r_i^y)$, and we set the lattice constant as unity. 
Similarly, the triple-$Q$ state is given by $\bm{Q}_1=(\pi, 0)$, $\bm{Q}_2 = (0, \pi)$, and $\bm{Q}_3 = (\pi, \pi)$ on the square-to-triangular lattice in Fig.~\ref{Fig:multipleQ_ponti_squ_tri}(c). 
We note that, as shown in Figs.~\ref{Fig:multipleQ_ponti_squ_tri}(a) and \ref{Fig:multipleQ_ponti_squ_tri}(b), all the relative angles between neighboring moments are $\cos^{-1} (0) = 90^{\circ}$ for the double-$Q$ state and $\cos^{-1} (-1/3) \sim 109^{\circ}$ for the triple-$Q$ state.

\subsection{Kondo Hamiltonian}
\label{sec:Kondo Hamiltonian}

In order to investigate the influence of the noncollinear and noncoplanar magnetic ordering on the electronic structure, 
we here consider the coupling of the magnetic moments to noninteracting electrons through the local exchange coupling. 
A minimal model to describe the situation is the Kondo lattice model, whose Hamiltonian is given by 
\begin{align}
\label{eq:Kondo_lattice_Ham}
\mathcal{H} = &-t \sum_{\langle i, j \rangle \sigma} (c^{\dagger}_{i \sigma} c_{j \sigma} +{\rm H.c.})  -t' \sum_{\langle \langle i, j \rangle \rangle \sigma} (c^{\dagger}_{i \sigma} c_{j \sigma} +{\rm H.c.}) \nonumber \\
&- J \sum_{i \sigma \sigma'} c_{i \sigma}^{\dagger} \bm{\sigma}_{\sigma \sigma'} c_{i \sigma'} \cdot \bm{S}_i, 
\end{align}
where $c_{i \sigma}^{\dagger}$ ($c_{i \sigma}$) is a creation (annihilation) operator of itinerant electrons at site $i$ and spin $\sigma$. 
The first term in Eq.~(\ref{eq:Kondo_lattice_Ham}) represents the hopping of conduction electrons on the nearest-neighbor bonds on the square lattice, while the second term is for the diagonal bonds introduced to connect the square lattice to the triangular lattice [see Fig.~\ref{Fig:multipleQ_ponti_squ_tri}(c)]. 
Note that $t'=0$ ($t'=t$) corresponds to the square (triangular) lattice. 
Hereafter, we take $t=1$ as an energy unit. 
The third term in Eq.~(\ref{eq:Kondo_lattice_Ham}) describes the exchange coupling between conduction and localized spins. 
Here, $\bm{\sigma}=(\sigma^x, \sigma^y, \sigma^z)$ is the vector of Pauli matrices, $\bm{S}_i$ is a localized spin at site $i$, and $J$ is the exchange coupling constant (the sign is irrelevant for the following mean-field type arguments).

By replacing $\bm{S}_i$ by Eqs.~(\ref{eq:doubleQ_definition}) and (\ref{eq:tripleQ_definition}), the Hamiltonian in Eq.~(\ref{eq:Kondo_lattice_Ham}) is written in the momentum representation, 
\begin{align}
\mathcal{H} = \sum_{\bm{k}\sigma} \varepsilon_{\bm{k}} c^{\dagger}_{\bm{k}\sigma} c_{\bm{k}\sigma} -
Jm \sum_{\bm{k}\sigma \sigma' \eta} 
c^{\dagger}_{\bm{k}\sigma} \sigma^\eta_{\sigma \sigma'} c_{\bm{k}+\bm{Q}_\eta \sigma'}, 
\end{align}
where $c^{\dagger}_{\bm{k}\sigma}$ and $c_{\bm{k}\sigma}$ are the Fourier transform of $c^{\dagger}_{i\sigma}$ and $c_{i \sigma}$, respectively; 
$\varepsilon_{\bm{k}}$ is the energy dispersion for free electrons, $\varepsilon_{\bm{k}}=-2t(\cos k_x + \cos k_y)-2 t' \cos (k_x - k_y)$. 
Here, the sum of $\eta$ in the second term is taken for $\eta=1$ and $2$ for the double-$Q$ state, and $\eta=1$, $2$, and $3$ for the triple-$Q$ state. 
We set the normalization factor $m$ so that $|\langle \bm{S}_i \rangle|=1$: 
$m=1/\sqrt{2}$ and $1/\sqrt{3}$ for double-$Q$ and triple-$Q$ states, respectively. 

In the momentum representation, the Hamiltonian is divided into two irreducible parts as 
\begin{align}
\label{eq:Ham_total}
\mathcal{H}=\bm{c}_{I}^{\dagger}\tilde{\mathcal{H}} \bm{c}_{I} +
\bm{c}_{I\!I}^{\dagger}\tilde{\mathcal{H}} \bm{c}_{I\!I}, 
\end{align}
where 
\begin{align}
{\tilde{\mathcal{H}}} =\left( \begin{array}{cccc}
\epsilon_{\bm{k}} & \Delta & - i \Delta & \alpha \Delta\\
\Delta & \epsilon_{\bm{k}+\bm{Q}_1} & - \alpha \Delta & i \Delta \\
 i \Delta & -\alpha \Delta & \epsilon_{\bm{k}+\bm{Q}_2} & \Delta \\
\alpha \Delta & - i \Delta & \Delta & \epsilon_{\bm{k}+\bm{Q}_3} 
\end{array} \right). 
\label{effective_Hamiltonian}
\end{align}
Here, $\bm{c}_{I}^{\dagger} =$ ($c_{\bm{k}\uparrow}^{\dagger}$, $c_{\bm{k}+\bm{Q}_1 \downarrow}^{\dagger}$, $c_{\bm{k}+\bm{Q}_2 \downarrow}^{\dagger}$, $c_{\bm{k}+\bm{Q}_3 \uparrow}^{\dagger}$), 
$\bm{c}_{I\!I}^{\dagger} =$ ($c_{\bm{k}\downarrow}^{\dagger}$, $c_{\bm{k}+\bm{Q}_1 \uparrow}^{\dagger}$, $-c_{\bm{k}+\bm{Q}_2 \uparrow}^{\dagger}$, $-c_{\bm{k}+\bm{Q}_3 \downarrow}^{\dagger}$), and $\Delta=Jm$. 
We introduce the parameter $\alpha$ to connect the double-$Q$ and triple-$Q$ states: 
$\alpha=0$ for the pure double-$Q$ state and $\alpha=1$ for the pure triple-$Q$ state.   
We note that this Hamiltonian with $\alpha=1$ is 
formally common to the four-sublattice triple-$Q$ orders on 
the triangular and cubic lattices~\cite{Martin_PhysRevLett.101.156402,Hayami:PhysRevB.89.085124}. 
In the two-dimensional triangular lattice case, the triple-$Q$ magnetic order induces Chern insulating states~\cite{Martin_PhysRevLett.101.156402,Akagi_JPSJ.79.083711} [see Fig.~\ref{Fig:multipleQ_band_squ_tri}(b)],  
while, in the three-dimensional cubic lattice case, it leads to semimetallic states with massless Dirac dispersions~\cite{Hayami:PhysRevB.89.085124}.

\subsection{Electronic structure}
\label{sec:Electronic structure}

\begin{figure}[hbt!]
\begin{center}
\includegraphics[width=1.0 \hsize]{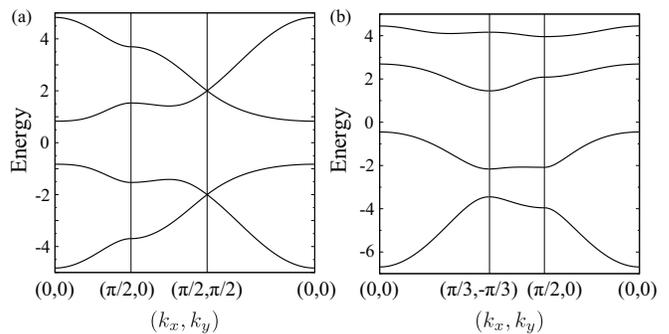} 
\caption{
\label{Fig:multipleQ_band_squ_tri}
Energy dispersions of (a) the noncollinear double-$Q$ state on the square lattice ($t'=0$) and (b) the noncoplanar triple-$Q$ state on the triangular lattice ($t'=1$) at $J=2$, shown along the symmetric lines in the magnetic Brillouin zone. 
In (a), two-dimensional massless Dirac nodes appear at $(\pi/2,\pi/2)$ at 1/4 and 3/4 fillings. 
Meanwhile, an energy gap opens at 1/4 and 3/4 fillings in (b). 
}
\end{center}
\end{figure}

The energy dispersion of the Hamiltonian in Eq.~(\ref{eq:Ham_total}) is shown in Fig.~\ref{Fig:multipleQ_band_squ_tri}(a) at $t'=0$, $J=2$ ($\Delta=\sqrt{2}$), and $\alpha=0$, which 
corresponds to the double-$Q$ state on the square lattice. 
All the bands are doubly degenerate; the degeneracy comes from the fact that $\bm{c}_I$ and $\bm{c}_{I\!I}$ are related by a combination of lattice translation and spin rotation, which leaves $\tilde{\mathcal{H}}$ unchanged. 
In Fig.~\ref{Fig:multipleQ_band_squ_tri}(a), a peculiar structure is found near the $(\pi/2,\pi/2)$ point; 
the band dispersions are linearly dependent on $\bm{k}$ and cross with each other at the $(\pi/2,\pi/2)$ point, resulting in two-dimensional cone-like structures. 
This is a signature of two-dimensional massless Dirac electrons appearing at 1/4 and 3/4 fillings of conduction electrons~\cite{affleck1988large,Yamanaka_PhysRevLett.81.5604,Agterberg_PhysRevB.62.13816,Sajeev_PhysRevLett.71.3343,Sajeev_PhysRevB.51.381}. 
In fact, by expanding the Hamiltonian around the $(\pi/2,\pi/2)$ point and performing the unitary transformations, the Dirac-type equation is obtained: 
\begin{align}
\label{eq:2D_Dirac_square}
\tilde{\cal H} \simeq \tilde{\mathcal{H}}_{\pm}^{{\rm 2D}} = \pm \sqrt{2} \Delta \sigma_0 \pm \sqrt{2}t(\kappa_x \sigma_z + \kappa_y \sigma_x), 
\end{align}
where $\sigma_0$ is the unit matrix and $\bm{\kappa}=(\kappa_x, \kappa_y)$ is the wave vector measured from $(\pi/2,\pi/2)$. 

On the other hand, the energy dispersion for the triple-$Q$ state on the triangular lattice is shown in Fig.~\ref{Fig:multipleQ_band_squ_tri}(b): the result is calculated at $t'=1$, $J=2$ ($\Delta=2/\sqrt{3}$), and $\alpha=1$ in Eq.~(\ref{effective_Hamiltonian}). 
In the triple-$Q$ state, the system becomes insulating at 1/4 and 3/4 fillings. 
The opening of the energy gap is due to the combination of the diagonal hopping $t'$ and the triple-$Q$ magnetic order parameter $\alpha\Delta$. 
In fact, the energy gap does not open when either $t'=0$ or $\alpha=0$.  
The insulating states at 1/4 and 3/4 fillings are topologically-nontrivial Chern insulators~\cite{Martin_PhysRevLett.101.156402,Akagi_JPSJ.79.083711}. 
Consequently, they exhibit the quantization of the Hall conductivity and gapless chiral edge states, as discussed below.

\begin{figure}[hbt!]
\begin{center}
\includegraphics[width=1.0 \hsize]{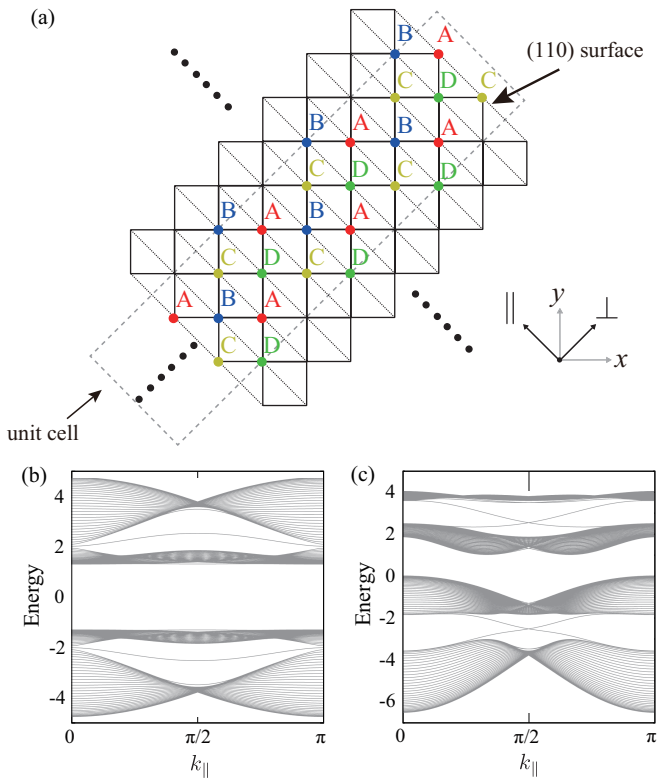} 
\caption{
\label{Fig:multipleQ_surface_squ_tri}
(Color online) (a) Schematic picture of the system with the (110) edges. 
The sites in the dashed box represent the unit cell used for the calculations of the edge states. 
The number of sites in the unit cell is 130.
The relation between ($k_x,k_y$) and ($k_\perp,k_\parallel$) is also shown. 
Energy dispersions for the systems at $J=2$ for (b) the double-$Q$ state on the square lattice ($t'=0$ and $\alpha=0$) and (c) the triple-$Q$ state on the triangular lattice ($t'=1$ and $\alpha=1$). 
}
\end{center}
\end{figure}

Let us discuss the electronic state in the multiple-$Q$ states more closely, with emphasis on the peculiar edge states associated with their topological nature. 
We here consider the system with the (110) open edges, in which both edges consist of A and C sublattice sites, as shown in Fig.~\ref{Fig:multipleQ_surface_squ_tri}(a)~\cite{comment_edge}. 
In the (1$\bar1$0) direction, we adopt the periodic boundary condition. 

Figure~\ref{Fig:multipleQ_surface_squ_tri}(b) shows the band dispersions of the system with the (110) edges for the double-$Q$ state on the square lattice. 
In this case, the Dirac nodes at the $(\pi/2,\pi/2)$ point in the bulk system are projected onto $k_{\parallel}=0$ and $\pi$. 
See Fig.~\ref{Fig:multipleQ_surface_squ_tri}(a) for the relation between ($k_x,k_y$) and ($k_\perp,k_\parallel$). 
Interestingly, there appear bands connecting the Dirac nodes for both 1/4 and 3/4 fillings. 
These are the chiral edge states appearing in the two-dimensional Dirac semimetal. 
The emergence of the bands connecting the Dirac nodes is similar to the case in zigzag-edged graphene nanoribbons; the bands, however, are dispersive in the present system, while they have no $k_\parallel$ dependence in the nanoribbons~\cite{Fujita_JPSJ.65.1920,Nakada_PhysRevB.54.17954,Hatsugai_PhysRevB.74.205414}. 

On the other hand, the triple-$Q$ state on the triangular lattice shows gapless chiral edge states traversing the gaps, as shown in Fig.~\ref{Fig:multipleQ_surface_squ_tri}(c). 
They have crossing points at $k_\parallel = \pi/2$ with linear dispersions. 
These are the gapless edge states emergent in the Chern insulating states, resulting in the nonzero quantized Hall conductivity~\cite{Martin_PhysRevLett.101.156402,Akagi_JPSJ.79.083711}. 

\subsection{Topological phase transition between the double-$Q$ and triple-$Q$ states}
\label{sec:Topological phase transition between square and triangular lattices}

\begin{figure}[hbt!]
\begin{center}
\includegraphics[width=1.0 \hsize]{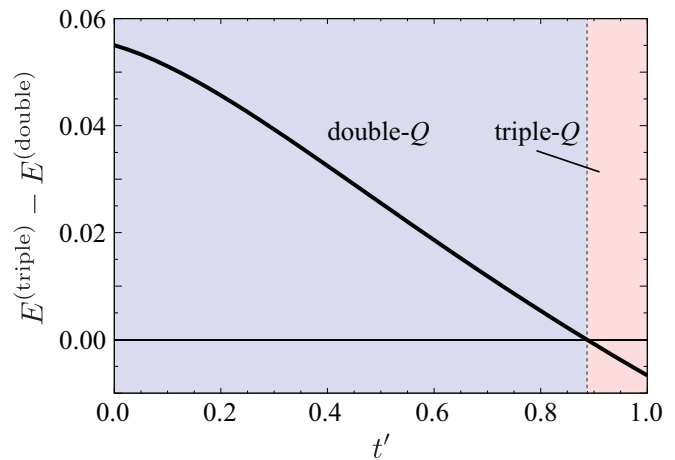} 
\caption{
\label{Fig:energy_multipleQ_graph}
(Color online) 
$t'$ dependence of the energy difference between the noncollinear double-$Q$ and noncoplanar triple-$Q$ states at 1/4 filling and $J=2$. 
}
\end{center}
\end{figure}

We here consider the stability of the multiple-$Q$ states from the energetic point of view. 
The noncollinear double-$Q$ state on the square lattice ($t'=0$) and the noncoplanar triple-$Q$ state on the triangular lattice ($t'=1$) were shown to appear at 1/4 filling in the Kondo lattice model by the Monte Carlo simulation combined with the exact diagonalization~\cite{Agterberg_PhysRevB.62.13816,Kato:PhysRevLett.105.266405}. 
Furthermore, the instability toward these multiple-$Q$ orders was explained by the perturbation in the weak coupling limit~\cite{Akagi_PhysRevLett.108.096401,Hayami_PhysRevB.90.060402}. 
Hence, the double-$Q$ and triple-$Q$ states are considered to be realized in the ground state, at least, at $t'=0$ and $t'=1$, respectively. 

Then, let us discuss how these two multiple-$Q$ states are connected with each other when we modify the lattice geometry by changing $t'$ from $0$ to $1$. 
We here simply compare the energy for the double-$Q$ and triple-$Q$ states while changing $t'$. 
Figure~\ref{Fig:energy_multipleQ_graph} shows the energy difference between the double-$Q$ state, $E^{{\rm (double)}}$, and the triple-$Q$ state, $E^{{\rm (triple)}}$, at $J=2$; $E^{{\rm (double)}}$ and $E^{{\rm (triple)}}$ are calculated for $\alpha=0$ and $1$, respectively.
When $t'=0$, the energy in the double-$Q$ state is lower than that in the triple-$Q$ state. 
As increasing $t'$, the energy difference becomes smaller, and finally, the energy for the triple-$Q$ state becomes lower for $t' \gtrsim 0.89$. 
Namely, the result indicates that a topological phase transition is expected between the Dirac semimetal and the Chern insulator at $t' \sim 0.89$.  
In the present simple energy comparison, however, 
we ignore a modulation of localized moments from the perfectly ordered double-$Q$ and triple-$Q$ states. 
We will examine whether this topological transition takes place for the periodic Anderson model, in which modulations of angles and lengths of the local moments are allowed.

\section{Stability of multiple-$Q$ states and topological phase transition}
\label{sec:Stability of multiple-Q states}

In this section, we discuss the stability of multiple-$Q$ states in a systematic way on the square-to-triangular lattice. 
We adopt a fundamental model for describing the coupling between conduction and localized electrons, the periodic Anderson model. 
After showing the Hamiltonian and the calculation method in Sec.~\ref{sec:Model and method}, we first present the ground-state phase diagram on the square and triangular lattices in Secs.~\ref{sec:Square Lattice} and \ref{sec:Triangular lattice}, respectively. 
We then discuss the phase transition between the noncollinear double-$Q$ and noncoplanar triple-$Q$ states on the square-to-triangular lattice in Sec.~\ref{sec:Connection between triangular and square lattices}. 
We also show the change of the edge states in the band structure associated with the phase transition. 

\subsection{Model and method}
\label{sec:Model and method}

The Hamiltonian for the periodic Anderson model is written as 
\begin{align}
\label{eq:PAM_Ham}
\mathcal{H} 
=& -t \sum_{\langle i,j\rangle \sigma} 
(c^{\dagger}_{i \sigma} c_{j \sigma} + {\rm H.c.})
-t' \sum_{\langle \langle i,j\rangle \rangle \sigma} 
(c^{\dagger}_{i \sigma} c_{j \sigma} + {\rm H.c.}) \nonumber\\
&- V \sum_{i \sigma}
(c^{\dagger}_{i \sigma} f_{i \sigma} + {\rm H.c.})
+ U \sum_i n_{i \uparrow}^f n_{i \downarrow}^f + E_0 \sum_{i \sigma} n_{i \sigma}^f, 
\end{align}
where $f^{\dagger}_{i \sigma}$ ($f_{i \sigma}$) is the creation (annihilation) operator of localized $f$ electrons with spin $\sigma$ at site $i$, and $n_{i \sigma}^f = f_{i \sigma}^{\dagger} f_{i\sigma}$. 
The first and second terms represent the kinetic energy of conduction $c$ electrons as in Eq.~(\ref{eq:Kondo_lattice_Ham}), the third term the on-site hybridization between $c$ and $f$ electrons, the fourth term the on-site Coulomb interaction for $f$ electrons, and the fifth term the atomic energy of $f$ electrons. 
The periodic Anderson model is reduced to the Kondo lattice model in Eq.~(\ref{eq:Kondo_lattice_Ham}) in the large $U$ limit with one $f$ electron per site; 
$f$ electrons give localized moments, which couple with conduction electrons via the Kondo coupling $J \propto V^2/U$. 
We focus on the commensurate filling, 
$n^{{\rm tot}} = (1/N) \sum_{i \sigma} \langle c_{i \sigma}^{\dagger} c_{i \sigma} + f_{i \sigma}^{\dagger} f_{i \sigma} \rangle =3/2$, which corresponds to the 1/4-filling case in the Kondo lattice model. 
Hereafter, we take $E_0=-4$.

In order to study the ground state of the model in Eq.~(\ref{eq:PAM_Ham}), we employ the Hartree-Fock approximation for the Coulomb $U$ term, which preserves the SU(2) symmetry of the interaction term as 
\begin{align}
n_{i\uparrow}^f n_{i\downarrow}^f 
&\sim 
n_{i\uparrow}^f \langle n_{i\downarrow}^f \rangle
+ \langle n_{i\uparrow}^f \rangle n_{i\downarrow}^f
- \langle n_{i\uparrow}^f \rangle \langle n_{i\downarrow}^f \rangle
-\langle f_{i \uparrow}^\dagger f_{i \downarrow} \rangle f_{i \downarrow}^\dagger f_{i \uparrow} \nonumber \\
&-f_{i \uparrow}^\dagger f_{i \downarrow} \langle f_{i \downarrow}^\dagger f_{i \uparrow} \rangle
+\langle f_{i \uparrow}^\dagger f_{i \downarrow} \rangle \langle f_{i \downarrow}^\dagger f_{i \uparrow} \rangle.  
\end{align}
Here, $\langle \cdots \rangle$ is the statistical average with respect to the one-body mean-field Hamiltonian. 
In the calculations, we adopt a $2\times2$-site unit cell to determine the phase diagram. 
We confirm that the phase diagram is not qualitatively altered in the calculations by using an enlarged $4 \times 4$-site unit cell.

\subsection{Square lattice}
\label{sec:Square Lattice}

\begin{figure}[hbt!]
\begin{center}
\includegraphics[width=1.0 \hsize]{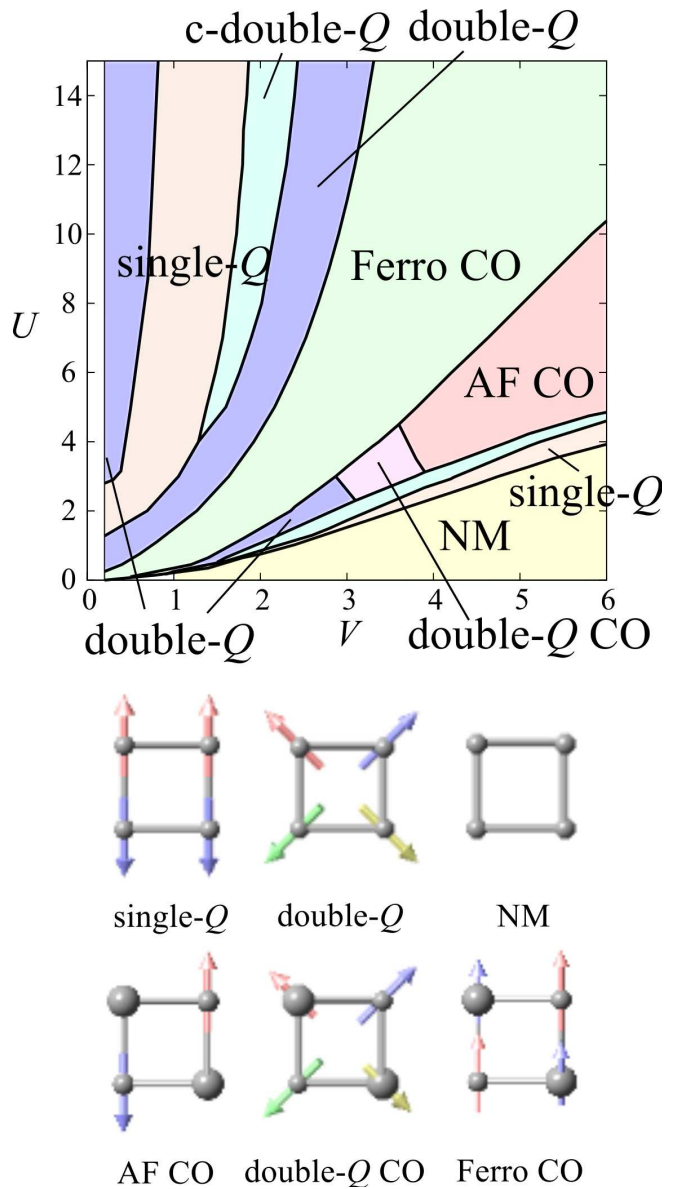} 
\caption{
\label{Fig:SL_U-V_phase_diagram_3}
(Color online) 
Ground-state phase diagram of the periodic Anderson model in Eq.~(\ref{eq:PAM_Ham}) on the square lattice at $n^{{\rm tot}} =3/2$ obtained by the mean-field calculations for $E_0=-4$ and $t'=0$. 
Schematic pictures of the ordering patterns in localized electrons are shown in the bottom panel. 
The sizes of the spheres denote local electron densities, and the arrows represent local spin moments. 
AF, Ferro, CO, and NM stand for antiferromagnetic (collinear N\'eel-type), ferromagnetic, charge-ordered, and nonmagnetic metallic states, respectively. 
Single-$Q$ corresponds to $\bm{Q}=(0,\pi)$ and double-$Q$ $(0,\pi)$, $(\pi,0)$. 
The c-double-$Q$ represents the double-$Q$ state with spin canting. 
}
\end{center}
\end{figure}

First, we discuss the stability of the double-$Q$ state with massless Dirac electrons by investigating the ground-state phase diagram of the periodic Anderson model in Eq.~(\ref{eq:PAM_Ham}) on the square lattice ($t'=0$). 
Figure~\ref{Fig:SL_U-V_phase_diagram_3} shows the result at $n^{{\rm tot}}= 3/2$ and $E_0=-4$ obtained by the mean-field approximation. 
The phase diagram is mainly divided into three regions; multiple-$Q$ region for large $U$ and small $V$ (for $U \gtrsim V^2$), charge ordered region for intermediate $U$ and $V$, and nonmagnetic region for small $U$ and large $V$ (for $V \gtrsim 2U$). 
Schematic pictures for each magnetic and charge ordered states of localized electrons are presented in the bottom of Fig.~\ref{Fig:SL_U-V_phase_diagram_3}. 

In the large $U$ and small $V$ region, we find the double-$Q$ state as one of the dominant magnetic phases. 
This phase accommodates the two-dimensional massless Dirac electrons in the electronic band structure as described in Sec.~\ref{sec:Multiple-Q states on lattices}. 
Dirac nodes also appear at $n^{{\rm tot}} = 1/2$, 5/2, and 7/2. 
We note that a similar double-$Q$ state was found by Monte Carlo simulation for the Kondo lattice model at 1/4 filling~\cite{Agterberg_PhysRevB.62.13816}, corresponding to $n^{\rm tot} = 3/2$ in the large $U$ limit with keeping $V^2/U$ at a nonzero constant. 
Interestingly, the double-$Q$ state in Fig.~\ref{Fig:SL_U-V_phase_diagram_3} appears around $U \propto 2V^2$. 

Other dominant instabilities in the phase diagram in Fig.~\ref{Fig:SL_U-V_phase_diagram_3} are the charge-ordered insulators. 
Interestingly, although the periodic Anderson model include neither bare off-site repulsive Coulomb interaction nor electron-phonon interactions, 
there are three charge-ordered phases in the intermediate $U$ and $V$ region: 
the ferromagnetic, N\'eel-type antiferromagnetic, and double-$Q$ charge-ordered phases. 
Among them, the charge ordered state with N\'eel-type antiferromagnetic state was found also in the Kondo lattice model~\cite{Misawa_PhysRevLett.110.246401}. 
Note that similar charge order instabilities were also pointed out in three-dimensional cubic systems~\cite{Zadeh_PhysRevB.55.R3332,hayami2013charge} and infinite dimensional systems~\cite{Majidi_2007charge,Otsuki_JPSJ.78.034719,Peters_PhysRevB.87.165133}. 
On the other hand, the double-$Q$ charge-ordered state has a noncollinear spin state; a similar but noncoplanar triple-$Q$ charge order was found on a cubic lattice~\cite{hayami2013charge}

\subsection{Triangular lattice}
\label{sec:Triangular lattice}

\begin{figure}[hbt!]
\begin{center}
\includegraphics[width=1.0 \hsize]{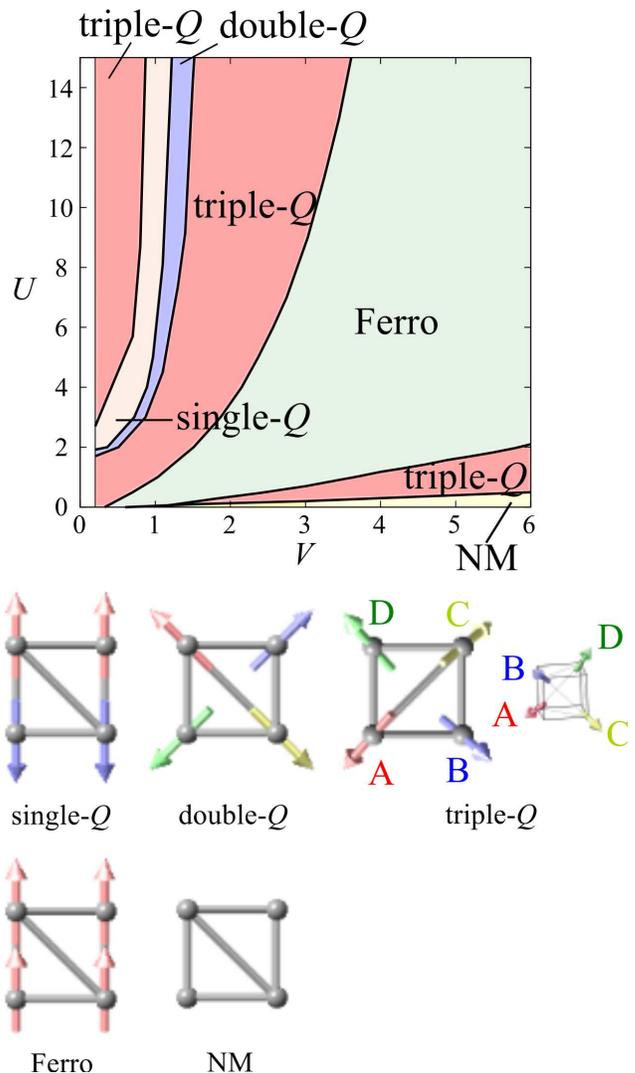} 
\caption{
\label{Fig:TL_n=1.5_U-V_phase_diagram2}
(Color online) 
Ground-state phase diagram of the periodic Anderson model in Eq.~(\ref{eq:PAM_Ham}) on the triangular lattice at $n^{{\rm tot}} =3/2$ obtained by the mean-field calculations for $E_0=-4$ and $t'=1$. 
Schematic pictures of the ordering patterns in localized electrons are shown in the bottom panel. 
The notations are similar to those in Fig.~\ref{Fig:SL_U-V_phase_diagram_3}. 
}
\end{center}
\end{figure}

Next, we discuss the stability of the multiple-$Q$ state on the triangular lattice ($t'=1$). 
Figure~\ref{Fig:TL_n=1.5_U-V_phase_diagram2} shows the ground-state phase diagram at $n^{{\rm tot}} =3/2$ and $E_0=-4$ obtained by the mean-field approximation. 
We find that the triple-$Q$ states are stabilized in the three regions; small $U$ and large $V$ (around $V \sim 6U$), intermediate $U$ and $V$ (around $U \sim 2V^2$), and large $U$ and small $V$ (for $U \gtrsim 3$ and $V \lesssim 1$) regions. 
We here focus on the former two, as the last one is replaced by other magnetically ordered phase when we include other additional interactions and hoppings in the model (not shown here). 

The stabilization mechanism of the triple-$Q$ state in the small $U$ and large $V$ region is attributed to the perfect nesting of the Fermi surface as explained below. 
For large $V$, the energy bands are split into the bonding and antibonding ones by the hybridization between $c$ and $f$ electrons, 
and the energetically-lower bonding band is partially filled at $n^{{\rm tot}}= 3/2$. 
The effective filling in the bonding band is $3/4$. 
As pointed out in Ref.~\onlinecite{Martin_PhysRevLett.101.156402}, the Fermi surface at 3/4 filling on the triangular lattice is perfectly nested, leading to the instability toward the triple-$Q$ order. 
This nesting instability is the origin of the triple-$Q$ order found in the small $U$ and large $V$ region. 

On the other hand, the stabilization mechanism is different for the triple-$Q$ state in the intermediate $U$ and $V$ region. 
In this case, the origin is understood by considering the large $U$ limit with keeping $V^2/U$ at a constant, where 
the periodic Anderson model at $n^{{\rm tot}} =3/2$ is reduced to the Kondo lattice model at 1/4 filling.  Indeed, the local density of localized electrons is approximately half filling in the large $U$ region in this triple-$Q$ state, which is considerably different from that in the triple-$Q$ state in the small $U$ and large $V$ region. 
In the Kondo lattice model, a similar triple-$Q$ order was found at and near 1/4 filling~\cite{Akagi_JPSJ.79.083711,Kato:PhysRevLett.105.266405}, and the origin was attributed to 
the ($d-2$)-dimensional instability of the Fermi surface~\cite{Akagi_PhysRevLett.108.096401, Hayami_PhysRevB.90.060402}. 
A similar mechanism may work in stabilizing the triple-$Q$ state in the intermediate $U$ and $V$ region in the periodic Anderson model. 

We note that both triple-$Q$ states are insulating and topologically nontrivial with nonzero Chern numbers. 
As in the Kondo lattice case~\cite{Martin_PhysRevLett.101.156402,Akagi_JPSJ.79.083711}, 
both states exhibit the quantum anomalous Hall effect.

\subsection{Connection between square and triangular lattices}
\label{sec:Connection between triangular and square lattices}

\begin{figure}[hbt!]
\begin{center}
\includegraphics[width=1.0 \hsize]{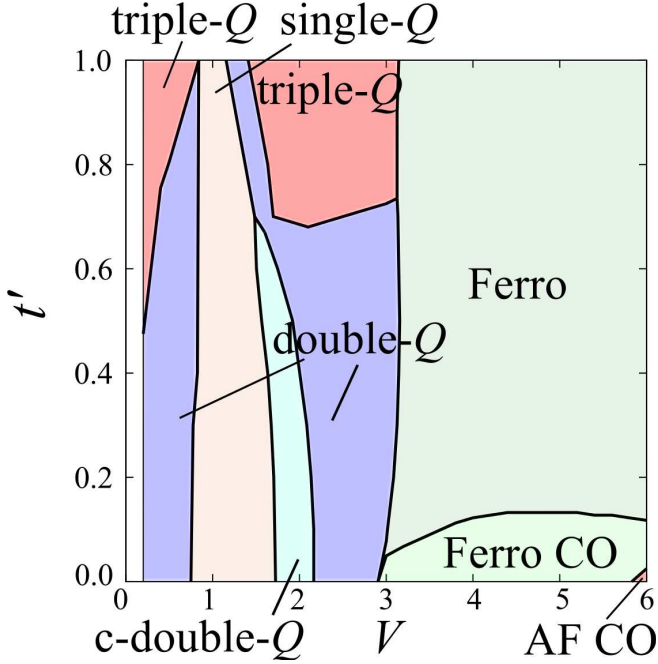} 
\caption{
\label{Fig:SL_TL-V_phase_diagram2}
(Color online) 
Ground-state phase diagram of the periodic Anderson model at $n^{{\rm tot}} =3/2$ for $E_0=-4$ and $U=10$. 
$t'=0$ corresponds to the isotropic square lattice system and $t'=1$ to the isotropic triangular lattice system. 
}
\end{center}
\end{figure}

In the previous two sections, we found that the double-$Q$ state on the square lattice and the triple-$Q$ state on the triangular lattice appear in the similar parameter region (along $U \sim 2V^2$) at $n^{{\rm tot}}= 3/2$. 
Now let us discuss the connection between the two states by changing $t'$ on the square-to-triangular lattice. 
Figure~\ref{Fig:SL_TL-V_phase_diagram2} shows the ground-state phase diagram at $U=10$ as a function of $V$ and $t'$. 
As described in Sec.~\ref{sec:Square Lattice}, there are two regions at $t'=0$: the multiple-$Q$ magnetically ordered region for small $V$ and the charge-ordered region for large $V$. 
These two regions exhibit distinct responses to the diagonal hopping $t'>0$ as described below. 

\begin{figure}[hbt!]
\begin{center}
\includegraphics[width=0.65 \hsize]{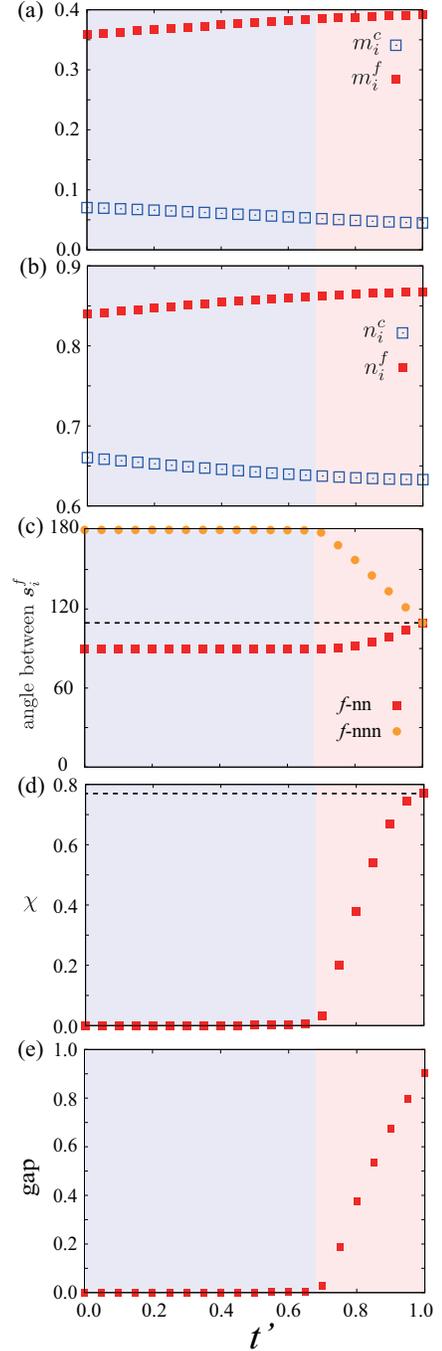} 
\caption{
\label{Fig:squ_tri_physical}
(Color online) 
$t'$ dependences of (a) the magnitude of local moments, (b) local electron densities, (c) relative angles between moments, (d) spin scalar chirality, and (e) energy gap. 
The data are taken at $n^{\rm tot}=3/2$, $V=2.5$, $U=10$, and $E_0 = -4$. 
In (c), $f$-nn and $f$-nnn mean the relative angles between nearest and next-nearest neighbor moments for $f$ electrons, respectively. 
In (c) and (d), the horizontal dashed line shows the values in the perfectly triple-$Q$ ordered states. 
}
\end{center}
\end{figure}

In the large $V$ region where the charge-ordered states are realized at $t'=0$, 
the system shows a continuous phase transition accompanied by the melting of charge order at a relatively small $t' \sim 0.1$. 
The resultant ferromagnetic state without charge disproportionation is extended to the triangular lattice case with $t'=1$. 

On the other hand, in the small $V$ region, the phase transition between multiple-$Q$ states occurs in a complicated manner. 
We here focus on the double-$Q$ state in the region of $2.2 \lesssim V \lesssim 2.9$.  
While increasing $t'$, 
as shown in Fig.~\ref{Fig:SL_TL-V_phase_diagram2}, the double-$Q$ state extends up to $t' \sim 0.7$, and shows a phase transition to the triple-$Q$ state, which continues to the triangular case with $t'=1$. 

The phase transition is a topological one between the double-$Q$ Dirac semimetal and the triple-$Q$ Chern insulator. 
Figure~\ref{Fig:squ_tri_physical} shows the changes of various physical quantities while changing $t'$ at $V=2.5$: the magnitude of local moments $m_i^{f(c)} = \sqrt{\langle s_{i,x}^{f(c)} \rangle^2 + \langle s_{i,y}^{f(c)} \rangle^2 + \langle s_{i,z}^{f(c)} \rangle^2}$ [$s_{i,\mu}^{f(c)}$ is the $\mu$ component of the spin operator for $f$($c$) electron], the local electron density $n_i^{f(c)}=\langle \sum_{\sigma} f_{i\sigma}^{\dagger}f_{i \sigma} (c_{i\sigma}^{\dagger}c_{i \sigma}) \rangle$, relative angles between nearest and next-nearest neighbor moments, the spin scalar chirality per plaquette for $f$ moments defined by $\chi=(1/N_{\alpha}) \sum_{\{i, j, k\} \in \alpha} \bm{s}^f_i \cdot (\bm{s}^f_j \times \bm{s}^f_k)/(|\bm{s}^f_i| |\bm{s}^f_j| |\bm{s}^f_k |)$ [($\bm{s}^f_i = (s^f_{i,x}, s^f_{i,y}, s^f_{i,z})$, $i$, $j$, and $k$ denote the sublattice index, and $\alpha$ labels each triangle plaquette; $N_{\alpha}$ represents the number of triangle plaquettes], and the energy gap. 

The magnitudes of the local spins and charge densities do not change significantly in the entire $t'$ region, as shown in Figs.~\ref{Fig:squ_tri_physical}(a) and \ref{Fig:squ_tri_physical}(b). 
On the other hand, the magnetic structure, scalar chirality, and energy gap show a drastic change in the phase transition, as shown in Figs.~\ref{Fig:squ_tri_physical}(c)-\ref{Fig:squ_tri_physical}(e). 
For $0 < t' \lesssim 0.7 $, the magnetic structure does not change so much from the double-$Q$ state at $t'=0$, and the energy gap remains zero. 
With further increasing $t'$ ($t' \gtrsim 0.7$), the $f$ moments are gradually canted in the perpendicular direction to the coplanar plane of the double-$Q$ order, as shown in Fig.~\ref{Fig:squ_tri_physical}(c). 
Accordingly, the spin scalar chirality takes a nonzero value and the energy gap opens, as shown in Figs.~\ref{Fig:squ_tri_physical}(d) and \ref{Fig:squ_tri_physical}(e). 
As $t' \to 1$, the magnetic order approaches to the pure triple-$Q$ one; the both relative angles between nearest-neighbor and next-nearest-neighbor moments become $\cos^{-1} (-1/3) \sim 109^{\circ}$, as shown in Fig.~\ref{Fig:squ_tri_physical}(c). 
At the same time, the value of the spin scalar chirality approaches $4/(3\sqrt{3})$ for the perfectly triple-$Q$ state. 
All these results coherently indicate that the phase transition occurs continuously between the double-$Q$ Dirac semimetallic state and the triple-$Q$ Chern insulating state at $t'\sim 0.7$. 

\begin{figure}[hbt!]
\begin{center}
\includegraphics[width=1.0 \hsize]{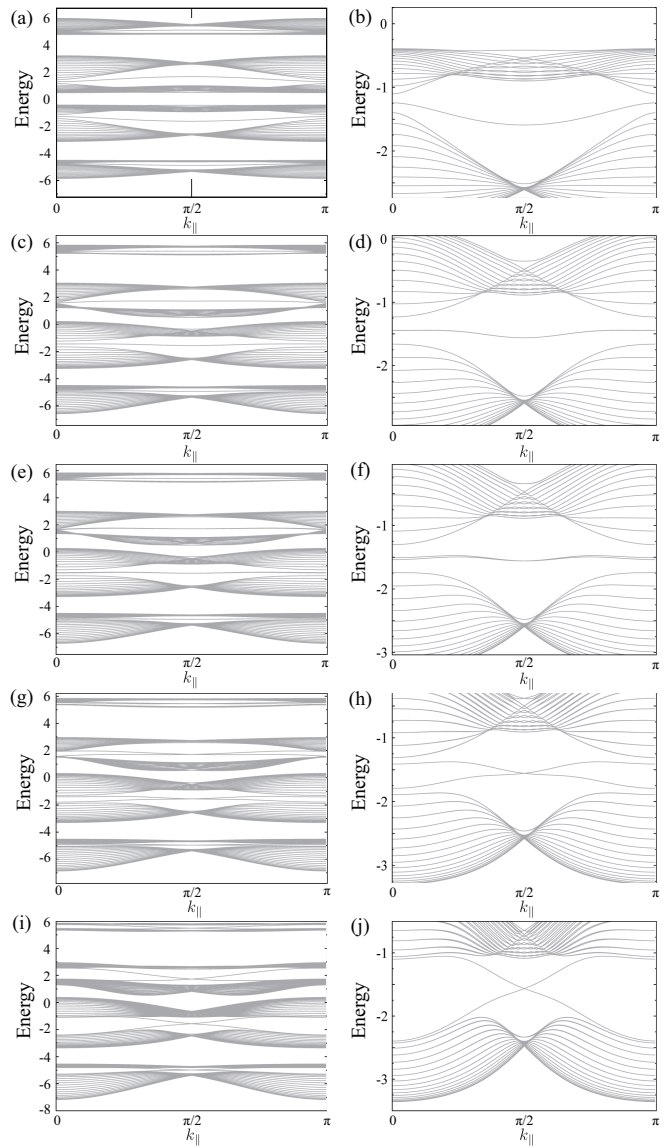} 
\caption{
\label{Fig:square_triangular_surface}
Energy dispersions for the system with the (110) edges at (a) $t'=0.0$, (c) $t'=0.6$, (e) $t'=0.7$, (g) $t'=0.8$, and (i) $t'=1.0$. 
The data are taken at $V=2.5$, $U=10$, $E_0=-4$. 
(b), (d), (f), (h), and (j) are enlarged figures of (a), (c), (e), (g), and (i) in the vicinity of the Fermi level at $n^{\rm tot}=3/2$, respectively. 
}
\end{center}
\end{figure}

Finally, we examine the topological transition between the multiple-$Q$ states from the viewpoint of the electronic structure by focusing on the edge states. 
We here consider the system with the (110) edges, as in Fig.~\ref{Fig:multipleQ_surface_squ_tri} in Sec.~\ref{sec:Electronic structure}. 
Figure~\ref{Fig:square_triangular_surface} shows the band structure of the system with the (110) edges at $V=2.5$ with varying $t'$. 
In Figs.~\ref{Fig:square_triangular_surface}(a) and \ref{Fig:square_triangular_surface}(b) corresponding to the square lattice case ($t'=0$), 
the edge states appear with connecting the Dirac nodes projected onto $k_\parallel=0$ and $\pi$. 
This is similar to the case in the Kondo lattice model as shown in Fig.~\ref{Fig:multipleQ_surface_squ_tri}(b). 
With increasing $t'$, the band structure does not show a significant change within the double-$Q$ state, as shown in Figs.~\ref{Fig:square_triangular_surface}(c) and \ref{Fig:square_triangular_surface}(d). 
Once the transition to the triple-$Q$ state occurs upon increasing $t'$ for $t' \gtrsim 0.7$, the dispersions show qualitatively different behavior. 
In the triple-$Q$ state, while the Dirac nodes for bulk disappear by opening an energy gap, the gapless dispersions from the edge states traverse the gap and cross with each other at $k_\parallel=\pi/2$, as shown in Figs.~\ref{Fig:square_triangular_surface}(g)-\ref{Fig:square_triangular_surface}(j). 
Namely, the degeneracy of edge states connecting the Dirac nodes in the double-$Q$ state is lifted according to the gap opening, and each edge state continuously develops to the chiral one in the Chern insulating state.

\section{Summary and Concluding Remarks}
\label{sec:Summary}
To summarize, we have investigated a phase transition between the noncollinear double-$Q$ state and the noncoplanar triple-$Q$ state in two-dimensional itinerant magnets. 
This is a topological transition between the Dirac semimetal and the Chern insulator. 
The result was shown for the Kondo lattice model and the periodic Anderson model on the square-to-triangular lattice. 
In particular, for the latter model, we explicitly showed that, by the mean-field approximation, the system exhibits a continuous phase transition accompanied by the growth of the spin scalar chirality and the energy gap while changing of lattice geometry. 
We examined the topological nature of the transition by the development of peculiar edge states. 

Our results provide a reference to further studies for the multiple-$Q$ states and the phase transitions between them. 
Especially, the phase transitions between double-$Q$, triple-$Q$, and paramagnetic states at finite temperatures will be stimulating. 
As a continuous symmetry cannot be broken at finite temperatures in two-dimensional systems if the interactions are short range~\cite{Mermin_PhysRevLett.17.1133}, the magnetic long-range orders will be destroyed in both the double-$Q$ and triple-$Q$ states at finite temperatures. 
Nevertheless, the scalar chirality in the triple-$Q$ state has a discrete symmetry, which can be broken at a finite temperature even in two dimensions; hence, it is expected that the system exhibits a scalar chiral spin-liquid phase at finite temperatures~\cite{Kato:PhysRevLett.105.266405}. 
On the other hand, in the double-$Q$ region, the vector chirality remains as an active continuous degree of freedom at finite temperatures, which might lead to an exotic phase transition, such as the Berezinskii-Kosterlitz-Thouless~\cite{berezinskii1971destruction,berezinskii1972destruction,kosterlitz1973ordering} and the $Z_2$ vortex transition~\cite{Kawamura_doi:10.1143/JPSJ.53.4138}. 
Thus, it is intriguing to clarify how the two multiple-$Q$ states develop in the finite-temperature region and how the system behaves around the quantum critical point between the two states. 
To investigate such exotic phenomena, it is necessary to perform an analysis beyond the mean-field approximation used in the present study. This is left for a future study. 

%%% Acknowledge %%%
\begin{acknowledgments}
The authors acknowledge Y. Akagi for fruitful discussions and careful reading of the manuscript. 
They also thank T. Misawa and Y. Yamaji for helpful comments. 
SH is supported by Grant-in-Aid for JSPS Fellows. 
This work was supported by Grants-in-Aid for Scientific Research (No.~24340076), the Strategic Programs for Innovative Research (SPIRE), MEXT, and the Computational Materials Science Initiative (CMSI), Japan. 
\end{acknowledgments}

\bibliographystyle{apsrev}
\bibliography{ref}

\begin{thebibliography}{57}
\bibitem[{com()}]{comment_edge}
\bibinfo{note}{When we cut one of the surfaces at B and D sublattice sites, a
  small gap opens at the Dirac cones. The gap magnitude decreases while
  increasing the system size along the $\perp$ direction and disappears in the
  bulk limit.}
\end{thebibliography}

\end{document}